\documentstyle[PASJadd]{PASJ95}

\begin{document}

\markboth{P.\ Khare and S.\ Ikeuchi}
{C and Si in QSO absorbers}

\title{The ionization and abundance of C and Si in QSO absorbers}

\author{Pushpa {\sc Khare}\\
{\it Physics Department, Utkal University, Bhubaneswar, 751004, India}\\
and\\
S. {\sc Ikeuchi}\\
{\it Department of Earth and Space Science, Osaka University,
Toyonaka, Osaka 560, Japan}}

\abst{
We have analyzed high resolution data of absorption lines of Si and C
in the absorption systems observed in the spectra of QSOs, in order to
study the ionization state and the overabundance of Si with respect to
C in the absorbers and also to study the change in these properties
with redshift.  No correlation is found between column density ratios
of Si IV to C IV of intervening systems and redshift.  The data do not
provide any evidence for an abrupt change in the values of the ratio at
any particular redshift unlike that for Lyman alpha forest clouds. We
have compared the observed ratios of column densities of Si II and Si
IV and of Si IV and C IV in different classes of absorption systems
with predictions of photo ionization models for different shapes of the
background radiation field.  Overabundance of Si over C can be ruled
out in several of the intervening systems for any shape of the
background radiation. For these systems we can also rule out any
contribution from the stellar sources to the background, which is then
entirely from the AGNs. No overabundance is needed in other intervening
systems if the radiation field from stellar sources contributes
significantly to the UV background. Overabundance is, however, 
present in Lyman alpha forest clouds at redshifts larger than 3 and 
in systems associated with the QSOs. For all the intervening systems a
minimum of 10 \% of the background is contributed by the AGNs.}

\kword{Galaxies: Quasars: absorption lines --- Galaxies: Abundances ---
Galaxies: intergalactic medium }

\maketitle
\thispagestyle{headings}

\section {Introduction}

Quasar absorption lines have proved to be extremely useful probes of
the  high redshift Universe. In recent years important information has
been obtained, among other things, about the chemical abundances in
galaxies at high redshifts. Evidence for evolution in chemical
abundances in these galaxies, the abundance increasing with decreasing
redshift,  has been obtained (i) directly, through abundance
determination in damped Lyman alpha systems (DLAS) at different
redshifts ( Pettini etal 1995; Lu etal 1996), and (ii) indirectly, from
the variation of the number of C IV systems and Lyman limit systems
(LLS) per unit redshift interval, per line of sight as a function of
redshift ( Steidel, Sargent {\&} Boksenberg, 1988; Khare {\&} Rana,
1993). The absorption line studies  have also given clues for
understanding the history of stellar nucleosynthesis (Lauroesch etal
1996). Lu etal (1996), through an analysis of the observed abundances
of 23 DLAS, have shown that the relative abundance patterns of several
elements are consistent with their formation in Type II supernovae. In
particular, they find evidence for [Si/Fe] $\simeq$ 0.4, which is very
similar to the overabundance of Si found in the Galactic halo stars.
Evidence for overabundance of Si with respect to C, by a factor of
three, has also been found in Lyman alpha forest clouds (LAFCs)
(Songaila {\&} Cowie, 1996; hereafter SC96) as well as in some
intervening and associated systems (Petitjean, Rauch {\&} Carswell
1994, Savaglio etal 1996).

The shape of the UV background radiation is important in deciding the
ionization balance of various elements in QSO absorbers. This radiation
most likely originates in AGNs, however, a significant contribution from
young star forming galaxies can not be ruled out (Bechtold etal 1987;
Bajtlik, Duncan $\&$ Ostriker, 1988, Madau {\&} Shull, 1996). The shape
of the background radiation not only depends on the relative
contribution from these two sources but is also decided by the
absorption by the material giving rise to QSO absorption lines.  Recent
observation of a significant Gunn-Peterson optical depth ($\tau\ge$1.7
for redshift $>$ 3.0) at the wavelength of the Lyman alpha line of He
II towards a QSO (Jackobsen etal 1994) indicates the presence of a
large break by factors $>$ 25 (Madau {\&} Meiksin, 1994) at the He II
ionization edge. Similar break was also found necessary in order to
explain the observations of column density of C and Si ions in the
Lyman alpha forest lines (SC96) as well as in some heavy element
systems (Savaglio etal 1996).  The shape of the background UV field due
to AGNs, at various redshifts, taking into account the absorption and
reemission from the QSO absorbers has been recently determined by
Haardt {\&} Madau (1996).

A lower He II optical depth ($\tau\simeq$1.0) at somewhat lower
redshift (z $\simeq$ 2.5) has been observed by Davidsen etal ( 1996)
towards HS1700+64, indicating a higher degree of ionization of He below
z=3.0. Evidence for the lowering of the optical depth of the Universe
to the He II ionizing photons below z=3.1 has also been obtained by
SC96, from the observations of the column density ratios of Si IV and C
IV in the LAFCs. They find an abrupt change in the values of this ratio
at z = 3.1, the ratio being higher at higher redshifts. As the
absorbers producing heavy element systems in the QSO spectra are also
most likely ionized by the intergalactic UV background (Srianand \&
Khare, 1995), it may be of interest to look for signatures of a change
in opacity of the Universe in the absorption line data of heavy element
systems.

A correlation with redshift of the ratio, of equivalent widths of lines
of Si IV ($\lambda$1393) and C IV ($\lambda$1548) in the heavy element
systems, observed at intermediate resolution, the ratio increasing with
increasing redshift, was found by Bergeron {\&} Ikeuchi (1990).  High
resolution data (FWHM $\le$ 23 km s$^{-1}$) for the column densities of
C IV and Si IV are now available for a number of QSOs. In this paper we
present the analysis of these data in order to understand the
ionization of these elements at different redshifts as well as to study
the overabundance of Si with respect to C in these systems. In section
2 we investigate the presence of a correlation between ratios of column
densities of Si IV and C IV and redshift. In section 3 we present the
results of photoionization models and compare these with the
observations. Conclusions are presented in section 4.

\section {Correlation Analysis }

We have collected from the literature (Cristiani etal 1995, Savaglio
etal 1996, Petitjean, Rauch {\&} Carswell 1994, Petitjean {\&} Bergeron
1994, Giallongo etal 1993, Fan {\&} Tytler 1994, Tripp, Lu {\&} Savage
1996, Prochaska {\&} Wolfe 1996, Wampler 1991, Wampler, Petitjean {\&}
Bergeron 1993) the column densities of C IV and Si IV observed in the
heavy element systems towards several QSOs observed with FWHM between 8
to 23 km s$^{-1}$. The total sample consists of 30 non-DLA 
intervening systems including 10 for which only upper
limits on the Si IV column density are available. In addition we have
23 components of intervening DLAS. 

We note that the data are rather inhomogeneous in the sense that the
observations have been made at somewhat different resolutions and with
different S/N values. This in principle can introduce an incompleteness
in the sample as the lower limit for detection of lines for different
QSOs may be different. This could affect the correlation analysis if
the incompleteness introduces a redshift dependence of the minimum
detectable column density, as a result, introducing an artificial
redshift dependence of the column densities in the sample. In order to
check this we performed Spearman rank correlation tests for redshift
dependence of column densities of C IV for the non-DLA and DLA
intervening systems separately. No such correlation was found, the
chance probabilities being 0.53 and 0.38 respectively. Similar
exercise for column densities of Si IV also rules out any correlation
with redshift, the chance probabilities being 0.84 and 0.74
respectively. We thus believe that our data, though inhomogeneous, are
not biased and can be used for the correlation analysis.

In Fig.1, we have plotted the ratios of column densities of Si IV and C
IV, R, and the upper limits on R as a function of redshift for both
categories of intervening absorption lines mentioned above. The non-DLA
intervening systems are believed to be produced by the
gas in the halos of absorbing galaxies and are most likely irradiated
by intergalactic UV background radiation (Srianand \& Khare, 1995).
Most DLAS have components with high as well as low ionization states.
These presumably arise in the halo and disk components of the absorbing
galaxy (Wolf etal 1996). Lu etal (1996) have pointed out that the Si IV
and C IV absorption profiles of DLAS always resemble each other and
have a different appearance from the low-ion absorption lines. They
suggest that the bulk of the high-ions could arise from the halo
clouds. The DLA components showing high-ions thus, very likely, belong
to the same population as that of the non-DLA intervening systems. We
can therefore combine the non-DLA intervening systems and the DLA
components showing high-ions together, while looking for the redshift
dependence of column density ratios of high-ions. 

Spearman rank correlation tests for  both categories of intervening
systems taken together as well as taken separately rule out any
correlation between R and z, the values of chance probability being
0.61, 0.229 and 0.213 for all intervening systems, non-DLA intervening
systems and DLA intervening systems respectively.  Generalized Kendall
test (Isobe, Feigelson \& Nelson, 1986) applied to the 53 values,
including upper limits, for all intervening systems and separately to
the 30 values, including upper limits, for the non-DLA intervening
systems gives probability of 0.684 and 0.174 respectively, which again
indicate the absence of any correlation. We also tried to study the
correlation for systems restricted to smaller redshift ranges. No
correlation was however seen. KS tests rule out any abrupt change in R
values at any particular redshift. Note that Spearman rank correlation
test applied to the values of R observed in LAFCs by SC96 gives chance
probability equal to 4.27$\times10^{-3}$, showing a good correlation.

\section {Ionization State of the Absorbers and the Background Field}    
 
For 14 systems ( 7 DLA intervening, 5 non-DLA intervening and 2
associated systems) the column density of Si II is also known. We can
thus try to investigate the shape of the UV radiation field as well as
the Si overabundance with respect to C in these systems. The neutral
hydrogen column density is not known for several of these systems.
However, as will be seen below, the exact value of this column density
is not very important. We have constructed a grid of photoionization
models using the code 'cloudy 84' written by Prof. G. Ferland, for
three values of neutral hydrogen column density, N$\rm_{H\;I}$, typical
of DLAS (10$^{20}$ cm$^{-2}$), of LLS (3.0$\times$10$^{17}$ cm$^{-2}$)
and of high ionization intervening systems ($\le 10 ^{16}$ cm$^{-2}$),
for different shapes of UV radiation field. Heavy element abundance was
assumed to be one thirtieth of the solar value, with solar values of
relative abundances of different chemical elements. The results, which
we present in the form of column density ratios, however, are not
sensitive to the heavy element abundances as well as the particle
density. The ratios are also independent of N$\rm_{H\; I}$ for
N$\rm_{H\; I}\;\le\; 10^{16}\; cm\;s^{-1}$. The ratios for these three
values of N$\rm_{H\; I}$ bracket the ion ratios for all the intervening
and associated systems considered here.

Heavy element absorbers have complex structures and it has been argued
that the lines of different ions may be produced in different regions
of the absorbers and also it is possible that a hot collisionally
ionized phase may exist in the absorbers (Giroux, Sutherland {\&}
Shull, 1994). However, here, we are restricting our analysis to high
resolution observations. We can therefore assume that lines produced in
individual clouds have been resolved and the the results of 'cloudy'
models, which take into account the radiation transfer inside a cloud,
can be applied to the column densities seen in individual clouds. Also
as we are restricting to the ions of Si II, Si IV, C II and C IV only,
the contributions from the collisionally ionized phase may not be very
important.

In Fig 2 we have plotted R vs. the column density ratios of Si II to Si
IV, S, for the three values of neutral hydrogen column densities
mentioned above. Fig 2a, 2b and  2c are for different shapes of the
background UV radiation field; these are (a) power law with a slope of
-1.5, which corresponds to typical unprocessed AGN spectra, (b) AGN
background filtered through the intervening galactic and intergalactic
absorbers as  given by Haardt {\&} Madau (1996, hereafter HM96) for the
redshift of 2.5 and (c) power law with slope of -1.5 and with a cutoff
at 4 Ryd, which will be the case if the He II ionization fronts of
different QSOs have not yet overlapped at the redshift of the absorbers
as suggested by SC96. We have also plotted the observed ratios with
their error bars. The observed values have large errors. These are
essentially a result of the fact that the lines are often saturated and
profile fitting analysis can yield acceptable fits to the observed
profiles over a range of column density values. However, as will be
seen below, definite conclusions regarding the overabundance and the
background can be drawn from the comparison of the observed values with
the results of the photoionization models. As seen from the figure the
shape of the spectrum makes significant difference to the lower values
of the ratios. Pure power law produces very low values of R for
S$\;<$0.1. Values of R for S$\;<$0.1 increase with increase in the
magnitude of the break at the He II ionization edge, however, the
maximum value of R is around 0.1 for infinite break. For larger values
of S, spectra of HM96 gives higher values of R compared to all other spectral
shapes. This is due to the decrement in spectra of HM96 short wards of hydrogen
ionization edge.

As the heavy element absorption systems are associated with galaxies,
it is possible that the radiation field incident on the absorbers may
get a significant contribution from the stellar sources. As the stellar
radiation field has very few photons beyond 3 Ryd, high values of R
are possible if the UV field is dominated by galactic contribution. We
have constructed photoionization models for clouds irradiated by
different proportions of galactic and UV background (HM96) fields.
Steidel (1995), from his study of a large sample of galaxies associated
with QSO absorption lines, finds these galaxies to be normal in the
sense of their star formation rates. We have therefore taken the shape
of the galactic field to be that given by Bruzal (1983). Pure galactic
background can not reproduce the observed ratios as it produces much
higher values of R, compared to the observed values. A contribution
to the radiation field from the intergalactic AGN background is necessary.
In Figure 2d we have plotted the column density ratios calculated for
the case when the ratio of the flux due to galactic radiation is 90{\%}
of the total flux at 1 Ryd, the rest 10\% being contributed by the AGN
background (HM96). The observed values of R for all the intervening
systems are lower than the results of this model. We can thus conclude
that a minimum of 10 \% (and possibly a much larger fraction, as will
be seen below) of the background radiation incident on the intervening
absorbers is contributed by the AGNs.

\subsection{Non-DLA intervening systems}

Assuming the shape of the radiation field incident on the absorbers
producing these lines to be that given by HM96, the observed ion ratios
for two of the systems, with redshifts 2.1 and 2.77, are consistent
with the results of the photoionization models. For other three
systems, one with redshifts 2.1 and two with redshifts 2.7, the
observed values of R, even after allowing for the errors are
considerably smaller than the model predictions. Adding galactic
radiation field to that of HM96 can only increase R, thereby increasing
the discrepancy. For other shapes of the background radiation the
observed ratios for four of the five systems are either consistent with
or are lower than the model values. For three of these systems we can
definitely rule out overabundance of Si w.r.t. C for any shape and
intensity of the background radiation. For these systems we can also
rule out any contribution of the galactic field to the radiation field.
This also holds for the remaining two systems if the shape of the
background is that given by HM96. Overabundance of Si by factors $>$ 
1.5 is necessary only for one system for other shapes of the background. 
Alternatively a small contribution from the stellar sources is required.

\subsection{DLA intervening systems}

For two of the systems, with redshifts 3.08 and 3.39, we can rule out
the overabundance of Si for any shape of the background. Between the
three shapes of the background radiation, the observed ion ratios for
other 3 sytems with redshifts between 1.7 and 3.38, are consistent with
the results of the photoionization models, without requiring any
overabundance of Si or any contribution from the galactic sources. For
the remaining two systems, with redshifts 2.84 and 3.39, the observed
values of R are higher than model results for all the three shapes of
the background. An overabundance of Si by factors $>$ 1.5 or a
significant contribution (Fig.2d) from the galactic sources is needed.

\subsection{Associated systems}

For the two associated systems with redshifts around 3.0, S is smaller
than 0.1 and R is larger than 0.3. Associated systems being close to
the QSO are expected to be ionized by pure power law radiation field.
Such a field, produces (Fig. 2a) smaller values of R. The
errors in the observed ratios are, however, too large and within those
error bars the ratios may be consistent with the model results.
However, though no overabundance is warranted by the observations due
to the large uncertainty in the observed values, one can not rule out
the possibility of Si being overabundant w.r.t. C by a large ($>$10)
factor in these systems.

\subsection{Lyman alpha forest clouds}

In Fig 3 we have plotted the observed ratios, R, vs. the ratio of
column densities of C II to C IV in Lyman alpha forest lines observed
by SC96. The theoretical results, assuming the UV background near the
Lyman alpha absorbers is purely from AGNs, for four spectral shapes (i)
power law with a slope of -1.0 (ii) power law with a slope of -1.5
(iii) power law with a slope of -1.5 with a cutoff at 4 Ryd and (iv)
field of HM96, are also plotted for N$\rm_{H\;I}\;=\;10^{16}$ km
s$^{-1}$. These results are independent of N$\rm_{H\;I}$ for
N$\rm_{H\;I}\;\le\;10^{16}$ km s$^{-1}$. No overabundance  
of Si is required for half of the LAFCs
which are mostly at redshifts smaller than 3.1
(SC96). The column densities for these systems are consistent 
with radiation 
field without any break at 4 Ryd. The amount of overabundance 
of Si needed for the other systems
(mostly at redshifts lager than 3.1) does depend on the shape of the
background and is smaller than a factor of 2 if a complete cutoff
beyond 4 Ryd is assumed. We thus agree with the conclusion of SC(96) 
that the data do indicate a break in the background at 4 Ryd at high 
redshifts indicating a higher He II opacity at these redshifts. 

\section{Summary}

We have analyzed high resolution observations of Si and C absorption
lines in the QSO spectra in order to study the ionization state of
absorbers and its change with redshift as well as to understand the
overabundance of Si w.r.t. C.

The column density ratios of Si IV to C IV in LAFCs, show a correlation
with redshift, as already noted by SC96. The observations of this ratio
in 30 non-DLA intervening absorbers, as well as in 23 intervening DLAS,
however, fail to show any correlation with redshift. The data do not
show any abrupt change in the ionic ratios at any particular redshift,
unlike the case of LAFCs noted by SC96. Thus there is no evidence for
the change in the opacity of the Universe beyond 4 Ryd from the column
density ratios of the heavy element line systems in the QSO spectra. It
may be argued that the radiation field incident on the heavy element
absorbers gets a significant contribution from local stellar sources,
thereby diluting the effect which is seen in the LAFCs. Our analysis of
ion ratios, presented in the previous section, argues against such a
possibility. Srianand and Khare (1995) have also presented several
arguments against such a possibility.
 
Observed column density ratios of Si IV to C IV and of Si II to Si IV
in several intervening and associated systems have been compared with
the results of photoionization models with different shapes of the
incident radiation. In spite of a large uncertainty in the observed
values, definite conclusions can be drawn about 
the overabundance. We find that for all the non-DLA intervening systems
the overabundance of Si can be ruled out if the shape of the
background is as given by HM96. For other shapes of the background also, the
overabundance can be ruled out for three of the five systems. The other
two systems may allow an overabundance by factors $>$ 1.5 if no
contribution from galactic sources to the background is assumed to be
present. For two of the DLA systems also, overabundance can be ruled
out for any shape of the background. Three other DLA systems are
consistent with the observations for shape of the background given by
HM96. The remaining two DLA systems, however, either require an
overabundance by factors $>$ 1.5 or a significant contribution from the
galactic radiation to the background. The possibility of overabundance
by factors $>$ 10 can not be ruled out for the associated systems. 
Lyman alpha forest clouds at high ($>$3) redshifts do indicate an 
overabundance of Si over C as well as higher opacity of the Universe 
to radiation beyond He II ionization edge at these redshifts.
\par
\vspace{1pc}\par
The authors are greatful to the Japan Society for Promotion of Science
and the Department of Science and Technology  (Government of India) for
sponsoring the collaboration. PK thanks the members of the Theoretical
Astrophysics group of the University of Osaka for warm hospitality and
R. Srianand for discussion.  This work was partially supported by a
grant (No. SP/S2/013/93) by the Department of Science and Technology,
Government of India.

\section*{References}
\small
\re Bahcall, J. N. et al 1993, ApJ, 87, 1
\re Bajtlik, S., Duncan R.C., Ostriker, J. P. 1988, ApJ., 327, 57
\re Bechtold, J., Weyman, R. J., Lin, Z. , Malkan, M. A. 1987, ApJ, 315, 180
\re Bergeron, J. {\&} Ikeuchi, S. 1990 A{\&}A, 235, 8
\re Bruzal, G. 1983, ApJS, 53, 497
\re Cristiani, S., D'Odorico, S., Fontana, A., Giallongo, E. , Savaglio, S. 1995, MNRAS, 273, 1016
\re Davidsen, A. F., Kriss, G. A. , Zheng, W. 1996, Nature, 380, 47
\re Fan, X. , Tytler, D., 1994, ApJS, 94, 17
\re Giallongo, E., Cristiani, S., Fontana, A. , Trevese, D. 1993, ApJ, 416, 137
\re Giroux, M. L., Sutherland, R. S. , Shull, J. M. 1994, ApJ, 435, L101
\re Haardt, F. , Madau, P. 1996, ApJ, 461, 20
\re Isobe, Feigelson , Nelson, 1986 ApJ, 306, 490
\re Jackobsen, P., Boksenberg, A., Deharveng, J. M.,
Greenfield, P., Jedrzejewski, R. , Paresce, F. 1994, Nature, 370,
35
\re Khare, P. , Rana, N. C. 1993, Journal of Astron {\&}
Astrophys, 14, 83
\re Lauroesch, J.T., Truran, J.W., Welty, D.E. , York,
D.G. 1996, PASP, 108, 641
\re Lu, L., Sargent, W.L.W., Barlow, T.A., Churchill C. W. ,
Vogt, S. S. 1996, ApJS, 107, 475
\re Madau, P. , Meiksin, A. 1994, ApJ, 433, L53
\re Madau, P. , Shull, J. M. 1996, ApJ, 457, 551
\re Petitjean, P. , Bergeron, J. 1994, A{\&}A, 283, 759
\re Petitjean, P., Rauch, M. , Carswell, R. F. 1994, A{\&}A, 291, 29 
\re Pettini, M., King, D. L., Smith, L. J. , Hunstead, R. W. 1995, 'QSO Absorption Lines' Ed: G. Meylan, (Springer), P 71
\re Prochaska, J. X. , Wolfe, A. M. 1996, preprint
\re Sargent, W. L. W., Boksenberg, A. , Steidel, C. C. 1988, ApJS, 68, 539
\re Savaglio, S., Cristiani, S., D'Odorico, S., Fontana, A., Giallongo, E. , Molaro, P. 1996, preprint
\re Songaila, A. , Cowie, L. L. 1996, AJ, 112, 335
\re Srianand, R. Khare, P. 1995 ApJ, 444, 643
\re Steidel, C. C., 1995, in 'QSO Absorption Lines' Ed: G. Meylan, (Springer), 139
\re Steidel, C. C., Sargent, W. L. W. , Boksenberg, A. 1988, ApJ, 333, L5
\re Storrie-Lombardi, L. J., McMahon, M. J., Irwin, M. J. ,
Hazard, C. 1996, ApJ, 468, 121
\re Tripp, T. M., Lu, L. , Savage, B. 1996, ApJS, 102, 239
\re Wampler, E. J., 1991, ApJ, 368, 40
\re Wampler, E. J., Petitjean, P. , Bergeron, J. 1993, AA, 273, 15 
\re Wolfe, A., Fan, X., Tytler, D., Vogt, S., Keane, M. J. , Lanzetta, K. M. 1994, ApJ, 435, 101
\break
\newpage
\centerline{\bf Figure Captions}

\noindent Fig 1: Plot of the ratio of column densities of Si IV and C
IV with redshift. Triangles and circles represent non-DLA and DLA
intervening systems respectively. Upper limits are indicated by T.\\

\noindent Fig. 2: Theoretical and observed column density ratios of Si
IV and C IV vs that of Si II and Si IV. The dotted line is for N$_{\rm
H\;I}\le10^{16}$ cm$^{-2}$, solid line is for N$_{\rm H\;I}=
3.0\times10^{17}$ cm$^{-2}$ and the dash-dotted line is for N$_{\rm
H\;I}= 10^{20}$ cm$^{-2}$. Circles correspond to the DLAS, squares
correspond to the associated systems and triangles correspond to the
non-DLA intervening systems. The shape of the UV continuum for Fig 2a,
2b and 2c is power law with slope =-1.5, Haardt {\&} Madau (96) spectra
for redshift of 2.5 and power law with slope =-1.5, with a cutoff at 4
Ryd, respectively. Fig 2d shows the results for a combined background
field due to galaxies (90$\%$ at 1 Ryd) and that given by Haardt {\&}
Madau (96) at the redshift of 2.5 (10$\%$ at 1 Ryd). Horizontal dashed
lines indicate the range of observed values.\\

\noindent Fig. 3: Theoretical and observed (SC96) column density ratios
of Si IV and C IV vs that of C II and C IV for Lyman alpha forest lines
for different shapes of UV background for N$_{\rm H\;I}\le10^{16}$
cm$^{-2}$. Solid line is for Haardt {\&} Madau (96) spectra for
redshift of 2.5; dashed line is for power law with slope of -1.0;
dotted line is for power law with slope of -1.5 and dash-dotted  line
is for power law with slope of -1.5 with a cutoff at 4 Ryd.\\
\break
\label{last}
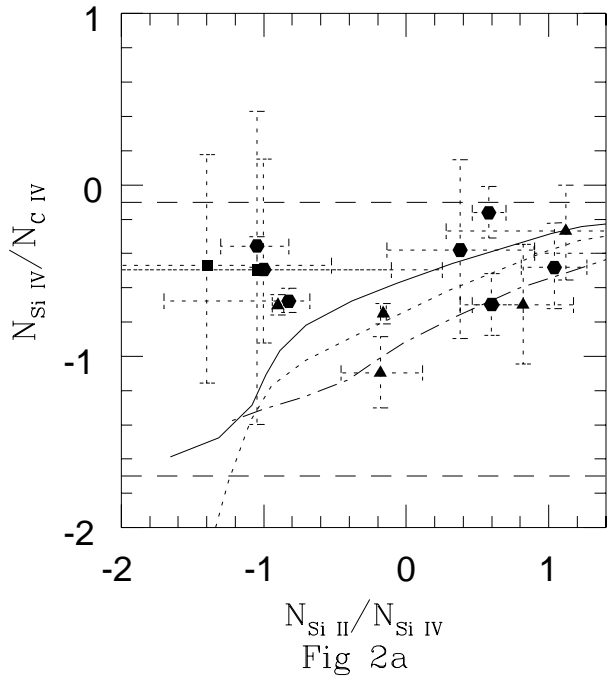
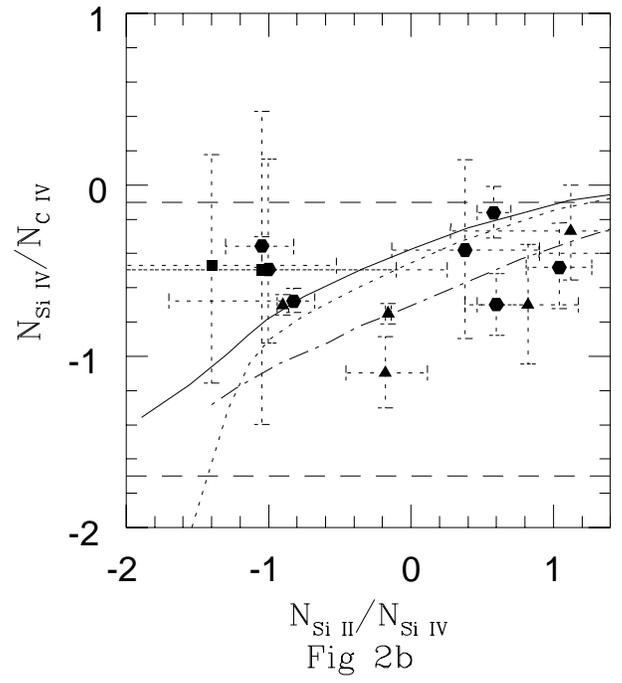
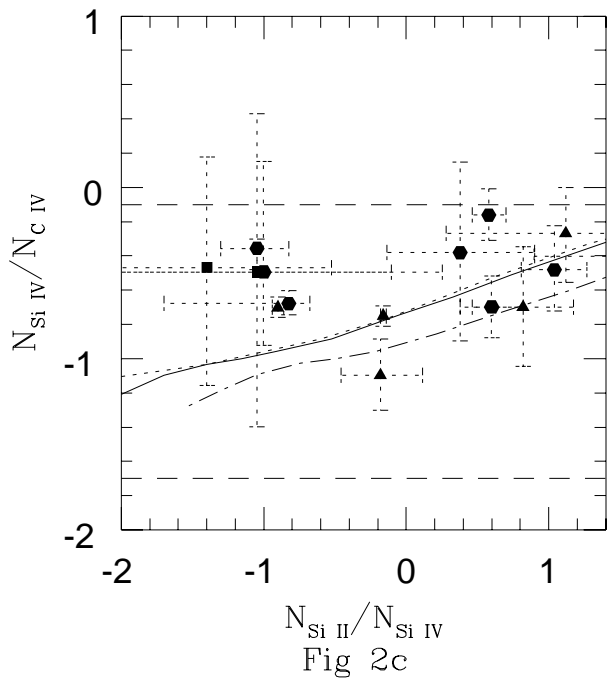
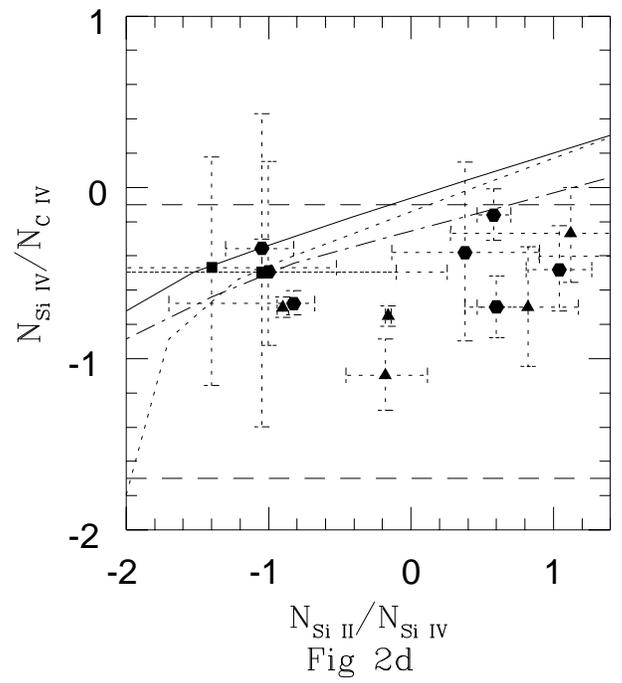
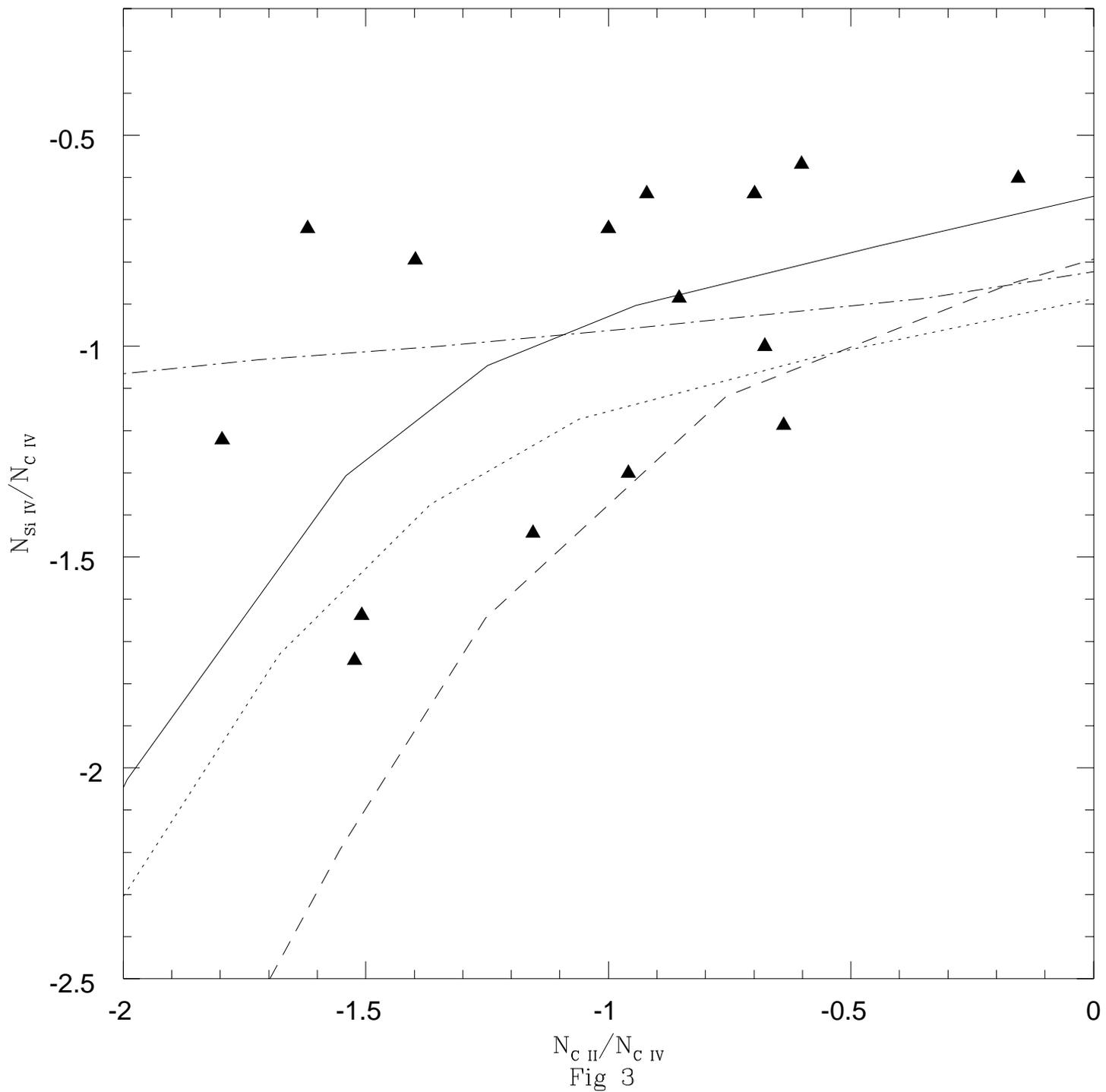
\end{document}